\documentclass[12pt,preprint]{aastex}

\slugcomment{Accepted for Publication ApJ Main Journal Nov 2002}

\shorttitle{Molecular and HI Segregation in Arp\,118}
\shortauthors{Appleton et al.}

\begin{document}

\title{Azimuthal and Kinematic Segregation of Neutral and Molecular Gas in Arp 118: The Yin-Yang Galaxy NGC 1144}

\author{P. N. Appleton}
\affil{SIRTF Science Center, MS 220-6, California Institute of Technology,  
Pasadena CA 91125}
\email{apple@ipac.caltech.edu}

\author{V. Charmandaris}
\affil{Astronomy Department, Cornell University, 106 Space Sciences Bldg, Ithaca NY 14853}
\email{vassilis@astro.cornell.edu}

\author{Yu Gao}
\affil{IPAC, MS 100-22, Infrared Processing \& Analysis Center, 
Caltech, Pasadena CA 91125 
\& Department of Astronomy, University of Massachusetts, LGRT-B 619E, 710 North Pleasant Street,Amherst, MA 01003-9305}
\email{gao@ipac.caltech.edu} 

\author{Tom Jarrett}
\affil{IPAC, MS 100-22, Infrared Processing \& Analysis Center, 
Caltech, Pasadena CA 91125}
\email{jarrett@ipac.caltech.edu}

\author{M. A. Bransford}
\affil{IBM, 3605 Hwy 52 North, Rochester, MN 55901}
\email{bransma@us.ibm.com}



\begin{abstract}

We present new high-resolution HI observations of the disk of the
collisional infrared luminous (L$_{\rm
IR}=2.2\times10^{11}$\,L$_{\sun}$) galaxy NGC\,1144, which reveal an
apparent large-scale azimuthal and kinematic segregation of neutral
hydrogen relative to the molecular gas distribution. Even among
violently collisional galaxies, the CO/HI asymmetry in NGC\,1144 is
unusual, both in the inner regions, and in the outer disk. We
suggest that we are observing Arp\,118 at a special moment, shortly
after a high-speed collision between NGC\,1144 and its elliptical
companion NGC\,1143. HI emission with an average molecular fraction
f$_{mol}$ $<$ 0.5 is observed on one side (NW) of the rotating disk of
NGC\,1144, while the other side (SE) is dominated by dense molecular
complexes in which f$_{mol}$ is almost unity. The interface region
between the warm-- and cool--cloud dominated regions, lies on a deep
spiral-like dust-lane which we identify as a shock-wave 
responsible for the relative shift in the dominance of HI and H$_2$
gas. A strong shock being fed by diffuse HI clouds with unusually
large ($>$ 400 km\,s$^{-1}$) rotational velocities can explain: 1)
the CO/HI asymmetries, 2) a large velocity jump (185 km\,s$^{-1}$)
across the arm as measured by HI absorption against a radio bright
continuum source which straddles the arm, and 3) the asymmetric
distribution of star formation and off-nuclear molecular gas resulting
from likely streaming motions associated with the strong-shock.  The
new results provide for the first time a coherent picture of
Arp\,118's many peculiarities, and underlines the potentially complex
changes in the gas-phase that can accompany large gravitational
perturbations of gas-rich galaxies.

\end{abstract}



\section{Introduction} 
   
It is now well known that galaxies in an intermediate stage of merger
conspire to create high concentrations of molecules in their nuclear
regions \citep{Sargent91,Scoville91,Sanders96,Bryant99}.  The creation
of these centrally concentrated molecular reservoirs may be quite
rapid because cases are rare in which molecular gas has not yet
settled to the center--although some are known
\citep{Lo97,Casoli99,Yun01, Wang01,Appleton02}. In a small subset of
cases, only one of the two participating galaxies contains nuclear CO
\citep{Dinh-V-Trung01,Evans02}, suggesting that the process of central
gas accumulation is complex--perhaps depending on the details of the
collision geometry, the initial gas content of the participants, or
other special circumstances \citep{Chen02,Yun01}.

There is little detailed knowledge of the effects of galaxy
collisions/mergers on the molecular fraction, f$_{mol} = M_{mol}/
(M_{mol}+M_{\rm HI})$, of the participating host galaxy disks. The
fate of any pre-existing neutral hydrogen in the disk of a collisional
galaxy might be expected to follow two different paths in major
merger.  HI in the outer galaxy is likely to become incorporated into
tidal tails, where it might survive the merger-at least temporarily
\citep{Hibbard96,Mihos01}. On the other hand, HI and pre-existing molecular
material in the inner galaxy is more likely to become involved in
inflows \citep{Noguchi88,Bekki94,Barnes96,Mihos96}.  \citet{Hibbard94}
speculated that the deficiency of HI in the inner regions of NGC\,7252
was a direct result of it being converted into other forms (ionized
gas, stars) through shock heating, a process that might lead to an
apparent segregation of molecular and HI gas in merging systems in
general \citep[e.g.][]{Duc97,Braine01,Yun01}. The high concentrations
of molecular gas in the nuclei of Luminous Infrared Galaxies (LIRGs)
and Ultra-luminous Infrared Galaxies (ULIRGs) \citep{Sanders96} could
be taken as evidence that some of the pre-existing HI is converted
into molecular form, but such ideas must be qualified by the fact that
most estimates of the molecular gas mass depend on a highly uncertain
conversion factor from the CO line flux to molecular hydrogen column
density \citep[e.g.][]{Sanders85,Scoville89,Downes98,Bryant99}.

Whatever the final outcome of any pre-existing gas in a merger, it is
likely that at least some of it will undergo a change of phase during
a violent encounter \citep{Gao99,GFN}, especially if it experiences
compression, or enters a high-pressure starburst region. Various
models governing such transformations have been outlined by
\citet{Jog92} specifically for starburst nuclei, and by
\citet{Elmegreen93} for a more general ISM environment.

In this paper we demonstrate that HI/CO segregation (defined here as a
separation of two regions with vastly different molecular fractions)
can occur outside the context of tidal bridges and tails: in this case
{\it within the rotating disk} of a violently interacting galaxy with
regular (albiet rapid) rotation.  We present new high resolution 21cm
HI observations of the interacting galaxy pair Arp\,118, comprising of
the disk galaxy NGC\,1144, and the elliptical/S0
companion NGC\,1143 (see Table 1).  An optical image of the system is
presented in Figure 1. It is believed that NGC\,1143 has reached its
current position on the sky after having collided with NGC\,1144 --
moving on a north-westerly path \citep{Lamb98}. NGC\,1144 has been
classified as a ring galaxy by \citet{Freeman74}, but its actual
morphology is far from simple \citep[see][]{Joy88}. Despite the
identification of peculiar arc-segments and ring-like structures
\citep{Hippelein89}, the H$\alpha$ velocity field shows remarkable
regularity \citep[see][ -- hereafter B99]{B99} suggesting that it
rotates in a counter-clockwise manner as projected onto the sky. The
most dramatic feature of its optical disk is a very extensive dark
dust-lane which originates east of the nucleus and extends outwards
curving to the north and west (see B99). If this is a trailing
feature, then the plane of the disk of NGC\,1144 is tilted with its
northwest side closer to the observer.

NGC\,1144 (a LIRG with L$_{\rm
IR}=2.2\times10^{11}$\,L$_{\sun}$) has a  Seyfert 2 spectrum
\citep{Osterbrock93}, and radiates 35\% of its $\lambda$10$\mu$m
luminosity from a giant extra-nuclear star forming complex situated to
the west of the active galactic nucleus \citep{Joy88}.  Other
morphological peculiarities are found in the radio
\citep{Condon90}. Aside from a weak radio emitting nucleus and
emission from the Mid-IR star forming knots, extended non-thermal
emission is seen along a prominent dust lane. Furthermore,
the brightest radio emission in Arp\,118 comes not from the nucleus,
but from a double radio source observed 8 arcsecs to the northeast of
the nucleus which straddles the dust lane. We will argue in Section~6, that
these sources are physically located in Arp\,118.

The peculiarities of NGC\,1144 do not end with its dramatically
different multi-wavelength appearance.  Kinematically, NGC\,1144 has
long been known to rotate extremely rapidly \citep{Hippelein89}
showing a spread in velocity of over 1200\,km\,s$^{-1}$ across its
disk (B99). \citet{Gao97} mapped the $^{12}$CO (1--0) emission in the
galaxy, and confirmed that the galaxy's molecular disk exhibits the
same rapid rotation as the ionized gas. Their high resolution CO maps
of NGC\,1144 showed that the molecular gas is {\it not concentrated at
the center of its disk}, but instead it is distributed asymmetrically
around it--a very unusual situation for an violently interacting
galaxy.  The strongest concentrations were found in the south-eastern
disk, the western star forming region and scattered clouds associated
with the parts of the northern segment of the outer ring. The rapid
rotation of the galaxy and the scattered ring was explained as being a
consequence of a slightly-off center, but nearly head-on collision
between NGC\,1143 and NGC\,1144 in the simulations of \citet{Lamb98},
leading to a spinning-up and partial expansion of the already massive
target galaxy. The models were able to reproduce the basic morphology
of the scattered ring, and one-armed spiral structure which dominates
the disk of NGC\,1144, though were less successful at modeling the
velocity field.

Our previous low-resolution VLA study of the HI in Arp\,118 (B99) was
made with the C-array resulting in maps with a synthesized Gaussian
beam with dimensions 21.2$\times$17.0 arcsecs (HPBW). Faint HI
(M$_{\rm HI}=7\times$10$^9$\,M$_{\sun}$) emission was detected over a
very narrow velocity range, and was concentrated towards the lowest
observed velocities in NGC\,1144. The HI seemed to anti-correlate with
the brightest CO emission, both in velocity-space and spatially,
although the spatial resolution was poor.  The interpretation of the
absence of HI over much of the velocity range of the CO emission was
not straight forward. Faint HI absorption was seen in two narrow
velocity bands within the velocity range in which CO (but not HI)
emission was detected. This hinted at the possibility that, within the
C-array beam, HI emission might be masked by deep absorption over the
same velocity range (roughly 800\,km\,s$^{-1}$). Although this was a
large range for HI absorption, HI profiles in other far-IR bright
galaxies are known to be affected by absorption \citep[see][]{Mirabel88}. We
predicted that if HI absorption had eaten away a large part of the
``normal'' disk HI profile, this would become obvious with higher
resolution observations--allowing the absorption and emission to be
easily separated spatially. This was the motivation for the new
observations made with the B-array of the VLA.

We have organized the paper as follows: Section 2 describes the
observations, Section 3 the HI observations, including the apparent HI/CO segregation and the HI absorption. In Sections 4 we discuss evidence for a
large-scale shock-wave in the disk of NGC\,1144. In Section 5 we discuss the possible origin of the HI/CO segregation, and present a
gas-flow scenario that seems to explain many of the observed
peculiarities of Arp\,118. In Section 6 we explore the importance of
the extremely luminous extra-nuclear radio sources which lie in the
disk of NGC\,1144. We present our conclusions in Section 7.  Following
B99, we will assume that H$_0$=80\,km\,s$^{-1}$\,Mpc$^{-1}$ and a
distance to Arp\,118 of 110\,Mpc.

\section{Observations}\label{sect_obs}

The VLA observations\footnote{The National Radio Astronomy Observatory
VLA is a facility of the National Science Foundation operated under
cooperative agreement by Associated Universities, Inc.} consisted of
an 8\,hr track made in B-array on June 23 1998, combined with a 4\,hr
track obtained previously in the C-array in 1996 (see B99). The new
data were flux calibrated using 3C48. Phase calibration was performed
by observing the nearby radio source 0320+052 (B1950) every 40 minutes
during the observations. The calibrated B-array visibility data (uv
data) were then merged with the calibrated uv data from the earlier
observations to form a new dataset using the AIPS routine UVCOMB. The
two separate IF band-passes were staggered such that two sets of 32
spectrometer channels overlapped, resulting in a velocity coverage of
2320\,km\,s$^{-1}$ (53 channels of width 193.5\,kHz = 43.8 km s$^{-1}$
in the rest frame of the galaxy) centered on V$_{helio}$ =
8721\,km\,s$^{-1}$. The spectrometer settings were identical to those
used in B99. The resulting channel maps had a synthesized HPBW of
6.2$\times$6.3 arcsecs. For some of the analysis we also created maps
with a larger synthesized beam of 8$\times$8 arcsecs and 15$\times$15
arcsecs, to improve our sensitivity to extended HI emission. We will
mainly present the 8$\times$8 maps here, resulting in $\sim$2.5 times
higher spatial resolution than the original B99 data. Continuum
subtraction was performed following the method described in B99.

As we shall show, HI emission and absorption was detected in a number
of separated bands in velocity space, with large numbers of empty
channels between these islands of emission.  As such, the creation of
an integrated HI map was more challenging than is usually the case
when observing HI emission spread continuously throughout the data
cube. Although the emission was easily seen in individual channels,
the usual methods of blanking and performing the traditional
``moment'' analysis over the entire observed velocity range created a
somewhat noisy result that did not match the quality of the individual
channels in which emission was detected.  As a result, we applied the
``moment'' analyses to groups of channels in which HI was detected,
and the final result was summed to create the integrated HI emission
map. Separate moment maps were created for those channels containing
absorption and emission. Because the emission and absorption occurs in
isolated regions of the data cube, we will not present iso-velocity
and velocity dispersion maps. However, the kinematics of the gas is
easily described in terms of the individual features in the cube.
 
\section{The HI Properties and Kinematics}

Figure 1 shows the integrated HI emission from the Arp\,118 system
superimposed on R-band images \citep[from][]{Hippelein89}.  Unlike our
previous measurements based on C-array observation alone, the new
observations resolve the HI clearly into emission regions which lie
in the north-western half of the irregular ring of NGC\,1144.
We confirm (Section 4) that the HI absorption is weak, and restricted
to the same narrow velocity ranges as reported in B99. A faint finger
of emission also extends outside the ring in the direction of the
elliptical companion NGC\,1143.  HI emission is absent to
the levels of detection (3-$\sigma$ upper limit is 0.85
mJy\,beam$^{-1}$channel$^{-1}$) from at least half of the disk of NGC\,1144,
including the southern and eastern portions of the ring, the major
star formation complexes in the south and western parts of the
disk. The observations confirm that lack of emission, rather than
emission eaten away by absorption, is responsible for the narrow HI
line-width of NGC\,1144.

Figure 2 shows the velocity field of the HI clouds derived from
inspection of the individual channel maps. For context, the systemic
velocities of NGC\,1144 is 8648$\pm$14\,km\,s$^{-1}$ and NGC\,1143 is
8459$\pm$30 km\,s$^{-1}$ (Keel 1996)\footnote{A number of different
values for the velocity of NGC\,1144 have been quoted in the
literature. The 2-D H$\alpha$ velocity field presented in B99 based on
Mt. Stromlo 2.3-m DBS spectroscopy shows that this velocity may not be
representative of the {\it disk gas} at that position, but is
characteristic of the velocity of the nucleus only. These
discrepancies may indicate that highly non-circular motions present in
the near-nuclear gas disk--a situation partly explained by our
proposed model for the gas flows presented in this paper.}. We have
labeled the emission clumps A through D for ease of identification.

Region A$'$ is the component of A that may be due to accreting gas
onto the companion galaxy NGC\,1143. Component A$'$ has a narrow
velocity width, and its mass is given in Table 1. The velocity of this
feature is lower by almost 200 km s$^{-1}$ from that of NGC\,1143, but
it may be gas which has been pulled from the outer parts of NGC\,1144 by
the violence of the collision. On the other hand, we cannot rule out
the possibility that it is disturbed gas in the outer regions of
NGC\,1144.

The clumpy emission at the north-western end of the outer ring (Region
A) occupies the lowest velocity channels (8227--8270\,km\,s$^{-1}$),
and it is in agreement with the lowest velocities seen in the optical
emission-lines in that region of the disk (B99).  At higher
velocities, a southern HI cloud complex B is seen at
8444--8531\,km\,s$^{-1}$ tracking the rotation of the galaxy, as does
region C to the north. These clouds have similar velocities to the
relatively regular (but very rapid) rotation of the ionized gas disk
of NGC\,1144 mapped out in B99. One exception is emission region D,
which has an extremely low velocity (V=8183\,km\,s$^{-1}$), and is
discrepant by approximately 900\,km\,s$^{-1}$ compared with its
expected velocity if it were to corotate with the inner ionized gas of
NGC\,1144 (No H$\alpha$ is observed in that direction). Since Region D
is seen in two adjacient channels at the 3 and 3.5--$\sigma$ level
respectively, we presume it is real, despite its peculiar
velocity. One explanation might be that the cloud is part of an
extension of a faint optical loop seen to the far-east of NGC\,1144 in
the red continuum, and even more faintly in H$\alpha$.  Given the huge
implied mass of NGC\,1144 (as implied from the rapid rotation speed
and the large velocity spread of 1200\,km\,s$^{-1}$ seen in the
ionized gas), this feature may be one of the few HI clouds seen in
these observations which is not part of the peculiar disk of
NGC\,1144.

\subsection{The ``Yin-Yang''-like HI and CO Segregation in Arp\,118}

In Figure 3 we present the HI integrated map combined with the high
resolution $^{12}$CO(1-0) map of the molecular gas seen in Arp\,118 by
\citet{Gao97}. The new observations reveal an apparent separation
between the CO and HI components, the CO emission being concentrated
mainly in the south-eastern quadrant of the galaxy, and the HI in the
north-western quadrant. Although there is some interleaving of the two
components far from the center of NGC\,1144, the HI and CO
emission appears strongly segregated azimuthally with respect to the
center--a highly unusual situation for a disk galaxy. It is this
segregation that leads to the narrow line width in the HI line, since
only part of the velocity field of the NGC\,1144 is being represented
in HI (see Figure 3 of B99 for a comparison with the CO profile).

In Figure 4 we present the integrated HI spectrum of the galaxy as
derived from the 8$\times$8 arcsecs (black histogram) and 15$\times$15
arcsecs channel maps (grey histogram), as well as the spectrum
obtained with the C-array only from B99 for comparison (open boxes).
We do not recover all the flux seen in the large-beam C-array
observations at the highest resolution (8$\times$8 arcsec beam), but
much more of the flux is seen in the intermediate-scale maps
(15$\times$15 arcsecs)--an indication that some of the gas is quite
extended on the 15--20 arcsec scale. In Table 1 we present tabulated
total masses of HI based on cubes made with small, intermediate and
large beams. In general, we refer to B99 for a full discussion of the
global properties of the galaxies. As the maps of B99 show, this
emission extends over the NW half of the galaxy, and exhibits the same
general anti-correlation with the CO as the more compact emission seen
here (See Figure 5 of B99 for example). Our observations (maps not
presented here for brevity) confirm the results of B99 for the
extended gas. In this paper we will concentrate on the results of the
observations with the highest resolution.

\subsection{The HI Absorption Lines}

Weak absorption features are seen in Figure 4, and they correspond
closely in velocity to those seen with lower signal-to-noise in the
C-array.  Of interest is the absorption System~2, which appears
slighly deeper in the current high-resolution observations than it did
in the earlier (B99) low-resolution observations. The reason for this
is also apparent in Figure 4. Faint emission, which is seen in the new
observations to the north of the nucleus in the outer ring at
V~=~8800\,km\,s$^{-1}$ (the highest velocity feature in Figure 2) is
observed at the {\it same velocity} as gas in the absorption
system--these two features cancelled each other out in B99.

With the exception of this minor effect, the new results demonstrate
that the absorption-line strength has not increased significantly
compared with the previously published lower-resolution observations.
In the earlier study we speculated that the absorption might be
significantly larger than was observed (the ``extreme absorber''
hypothesis--see B99), if the absorption extended over the same
velocity range as possible putative emission of similar strength
within the large beam. However, the current observations do not reveal
strong HI emission from the SE disk, nor is deep broad absorption
detected.  We can therefore rule out the peculiar shape of the HI
profile as being a result of deep broad HI absorption.

In Figure 5 we present integrated maps of the HI absorption covering
the velocity range of the two HI absorption features: System~1 at
8967--9099\,km\,s$^{-1}$, and the slightly deeper System~2, at
8749--8880\,km\,s$^{-1}$.  The two systems are separated by
$\sim$185\,km\,s$^{-1}$, and do not share the same spatial centroid
position. Instead they straddle the deep dust lane which crosses
through the southern component of the bright double radio source (see
Section~7), suggesting that the two components might be sampling the
velocity jump across a very powerful shock wave. We will return to
this topic in Section 4.

The observed and derived properties of the absorption lines are given
in Table 2. In addition to the kinematic properties of the absorption
features (columns 2 and 3), we also present the average and maximum
values of the 21cm line optical depth derived from the depth of the
line at each channel. The results were derived from the 6.2 $\times$
6.3 arcsec maps (see Section 2) to provide the highest discrimination
of the background continuum. Care was taken to ensure that the
continuum appropriate to the particular absorption system was used,
since Systems 1 and 2 have different centroids. The derived HI column
densities (Table 2: Column 6) assume the HI is uniformly distributed
across a source having the same angular dimensions as the VLA beam,
and as such represent strictly lower limits to the actual optical
depths.  More realistically in B99, despite the lower resolution, we
estimated the optical depth by assuming that the HI absorption was
seen against the compact double source, and used the fluxes for those
sources on the arcsec scale from \citet{Condon90}.  Using those
assumptions, the column density was found in each line to be somewhat
larger than in Table 2. Realistic values probably lie in between, and
hence we conclude that the HI column densities in the two lines are in
the range 5.5 $\times$ 10$^{20}$ to 2 $\times$ 10$^{21}$ atoms cm$^{-2}$.

In B99, we concluded that these column densities were consistent with
10--15 HI clouds of mass M$_{\rm HI} \sim 8~\times 10^{6}~$M$_{\odot}$ within the VLA
beam in order to avoid detecting the clouds in emission rather than
absorption. These estimates are in good agreement with the new
observations, if we make similar assumptions about the filling factors and
continuum radio source sizes. Interestingly, the above HI column
densities are high enough that it would be expected that the gas would
be mainly molecular. The transition from mainly atomic to mainly
molecular gas is believed to occur somewhere in the range of HI column
densities between $4 \times 10^{20}$ and $6.5 \times 10^{20}$ at cm$^{-2}$~
(3-5~M$_{\odot}$ pc$^{-2}$) \citep{Reach94, Hidaka02}. Hence the HI absorption
may be probing atmospheres of dense molecular condensations. Oddly, no
bright CO emission is detected from the position of the bright radio
source (see Figures 2 and 4). We will return to this point in Sections 6 \& 7. 

All things being equal, we can conclude from Table 2, that the HI
column density for System~2 is almost twice that of System~1, a fact
that is also consistent with the idea that the higher--velocity feature
is sampling down-stream gas from the dust-lane/shock front (see
Section 6).

\section{Evidence for a Global Shock-wave in the Disk of NGC\,1144}

In this section we will present new evidence that suggests that
NGC\,1144 contains an unusually powerful global shock-wave coincident
with a deep dust lane.  The shock-wave may be an important clue as the
mechanism for the apparent segregation of the HI and molecules in NGC
1144 (see Section 6)--inspiring the term ``Yin-Yang'' in the
title of this paper.

Figure 6a shows an archival R-band (F606) continuum image of the inner
portion of NGC\,1144 obtained with WFPC2 on the Hubble Space Telecsope
(HST). The dominant features of the disk are a central bright bulge
with fragmentary spiral arms, a massive star formation complex to the
west of the nucleus \citep[this is the same region which emits so much
mid-IR luminosity][]{Joy88}, and a prominent dust lane (indicated by
the arrows) which runs clockwise from a point southeast of the
nucleus, towards the northwest. A further loop of star-clusters is
also seen running from east to west along the southern extremities of
NGC\,1144, where it twists north to join the large western
star-forming complex.

What is the nature of the dust-lane in NGC\,1144 which seems so
striking at optical wavelengths? Radio continuum observations may
provide part of the answer. The optical morphology of the dust-lane is
more easily traced in Figure 6b, which shows the same HST image after
the application of an unsharp-mask. This helps to identify the high
spatial-frequency components in the image, showing the strength and
extent of the dust-lane, as well as allowing several shell-like
features to be perceived to the west of the nucleus. We have also
superimposed in this image the contours of the 20cm radio continuum of
NGC\,1144\footnote{The VLA radio continuum map of the A-array
observations were kindly supplied by J. Condon (NRAO) and were
presented in \citet{Condon90}.}. A close correspondence is seen
between the position of the dust lane and the radio emission which
seems to follow it all the way from its north-western tip down through
the bright elongated source \citep[this is resolved as a double by][]
{Condon90}, and through the twist of the dust-lane to the south and
west. The southern component of the double source (see arrow in
Figure~6b), lies suspiciously near the center ridge-line of the dust
lane, providing further evidence that they might be causally related.
Radio emission is also detected from the active nucleus and the
western star forming complex.

The association of fainter radio emission with the dust-lane is strong
evidence that the lane is a compression of the ISM. Similar radio
continuum enhancements were detected in the ring galaxy VII\,Zw\,466
\citep{Appleton99}, and are common in shock waves associated with
spiral arms \citep{Rohde99,Wielebinski01}.  Cosmic ray particles
radiate more efficiently when the magnetic field is compressed and
magnetic field irregularities are pushed closer together, trapping the
cosmic rays for longer in the disk \citep[see][]{Bicay90,Appleton99}.
Hence the radio observations support the idea that the dust-lane is a
shock wave. 

Further evidence that the dust-lane structure is purely hydrodynamic in
nature rather than a traditional spiral density wave, is supported by
the lack of association of an old stellar population with the
dust-lane/filament. We show in Figure 7 the H$\alpha$ image of the
galaxy superimposed with contours of the near-IR emission, as traced
by the sum of its J, H, and K--band 2MASS images. The
H$\alpha$ filament and dust lane have no counterpart in the
near-infrared continuum light, strongly suggesting that the feature is a
shock-wave in a purely gaseous disk with little detectable emission
from an underlying stellar component.

Finally, there is kinematic evidence linking the dust lane to a
powerful shock-wave. The HI absorption lines (Figures 4 and 5) show
two velocity components separated by 185\,km\,s$^{-1}$.  Could the two
components represent the up-stream and down-stream components of the
gas entering and leaving the shock? Most normal streaming motions
associated with spiral arms show quite modest velocity changes (10--20
\,km\,s$^{-1}$), so such a velocity jump across a shock would require
unusual circumstances. The two components straddle the dust-lane, the
higher velocity feature, System~1, is observed on the eastern side of
the dust lane, whereas System~2 is to the west. Given the extremely
rapid rotation of gas in NGC\,1144 (circular velocities of at least
400 km s$^{-1}$) it would be relatively easy to deflect gas passing
into the shock by an angle of $\sim$30 degrees, in order to bring a
substantial component of the rotational velocity into the radial
line-of-sight direction--thereby increasing its line-of-sight velocity
component.  Further hints of this may be correct come from the
kinematics of the CO clouds (clouds 9, 10 \& 11 of \citet{Gao97}) which
show much higher velocities east of the dust lane than is observed
along it. This is not unreasonable if the gas
streams almost tangentially to the dust-lane after crossing the shock.
If this is correct, then we would associate System~1 with neutral gas
reforming down-stream of the shock, and System~2 with the upstream
(unshocked) gas. In this case, the lower implied HI column of System~1
would result from this gas being depleted through ionization and/or
molecular cloud formation \cite[see][]{Elmegreen93}.

\section{The Origin of the Apparent Segregation of Molecular/HI gas}  

The asymmetric azimuthal distribution and apparent segregation of HI
and CO emission in NGC\,1144 is very unusual when compared with both
normal and violently collisional galaxies. Normal galaxies (whether
barred or unbarred) generally have symmetrically disposed CO
distributions with gas tending to centrate towards the center. For
example, in a subset of the BIMA/SONG survey of nearby
galaxies \cite[e. g.][]{Regan01}, only one out of 15 galaxies showed strong
asymmetries in the CO distributions--and even in that case (NGC\,3627)
the asymmetry was in the length of one spiral arm relative to
another. None of these galaxies show the peculiar lop-sided
CO distribution of NGC\,1144. Massive CO clouds
extending asymmetrically from the inner regions to the outermost
extent of the disk are apparently very rare. Although HI asymmetries are more
common in the outer parts of normal galaxies \cite[see for example
the recent study by][]{Jog02}, these are almost always associated with
the faintest outer HI isophotes, the inner regions are usually less
asymmetric.

However, even in interacting or merging systems, HI or CO asymmetries
of the kind seen in NGC\,1144 are surprisingly rare. If the two
components are segregated, it is almost always  radial segregation with
strong CO emission at the center coincident with a
nuclear starburst, HI in the outer parts in obvious tidal tails \citep[e. g.][]{Duc97, Yun01}. Even in the rare case where azimuthal
asymmetry is seen in the outer HI component \citep[e. g. Mk 273 or NGC
3921][]{Hibbard96b,Hibbard96}, the HI is clearly part of a narrow, highly distinct
tail, and not part of a coherent rotating disk, as in the case of
NGC\,1144. 

Despite the differences between NGC\,1144 and normal galaxies, studies
of normal galaxies are not without relevance to this case.  Molecular/HI phase
transitions within normal disks have led to the concept of the
``molecular front'' pioneered by \citet{Sofue95} and \citet{Honma95}, and
based on the theoretical ideas of \citet{Elmegreen93}. Here,
an HI/H$_2$ phase-transition was used to explain
why many galaxies are dominated by molecules in the center, but change
suddenly to being HI--dominated further out.  According to this
picture, f$_{mol}$ is critically dependant on the pressure confining
the clouds P$_e$, the strength of the UV radiation field, and the
metallicity of the gas. In normal galaxies, these parameters decline
roughly exponentially with galactocentric radius, and provide a reasonable
description of the sudden change from molecules to HI in a normal
disk. In a collisional system like NGC\,1144, the conditions in the galaxy are likely
to be very different from the quiescent case. In particular, a
non-axisymmetric one-armed shock could provide the necessary
conditions for azimuthal CO/HI asymmetries.  

In Arp\,118, we have measured the value of $f_{mol} < 0.5 $ in the
bright HI regions to the west of the shock-wave, based on
(3--$\sigma$) upper limits of M$_{\rm H_2}$ $<$ 5.7 $\times$ 10$^8$
M$_{\odot}$ from the CO observations of \citet{Gao97}\footnote{The
limit is based on the ``standard' CO flux to H$_2$ mass conversion
factor of 3 $\times$ 10$^{20}$ [mol cm$^{-2}$]/[K km\,s$^{-1}$].}
(after scaling to the HI beam), and is $>$ 96\% in the region of the
bright molecular clouds, based on a similar HI upper limit of M$_{\rm
HI}$ $<$ 1.1 $\times$ 10$^8$ M$_{\odot}$ from this paper. These
results suggest that the CO and HI are not truly segregated, but that
the molecular fraction changes dramatically across the shock-wave.

To demonstrate a possible causal relationship between the possible
shock-wave and the CO/HI asymmetry, we show the distribution of CO and
HI emission with the position of the shock-wave superimposed in Figure
8. The position of the shock is assumed to be coincident with the
dust-lane/radio-continuum features of Figure 6. This demonstrates a
number of interesting facts: 1) The majority (but not all) of the HI
lies to the west of the dust-lane, 2) except at large galactocentric
distances, all the centroids of the dense molecular cloud lie to the
east of the dust-lane in the clouds associated with the northern half
of the galaxy, and 3) HI absorption is seen {\it east} of the dust
lane in a gap in the CO clouds. This gap (mentioned earlier) is
centered on the 50 mJy radio source. The gap suggests that some local
feed-back effects have disturbed the process of HI to H$_2$ conversion
by rapidly turning H$_2$ into high mass stars and supernovae
(Section~7).  This supports the idea that the radio source is
localized within Arp\,118 and is not a powerful background object.

The strong shock would naturally explain why the disk of NGC\,1144 is
mainly molecular down-stream of the shock, and mainly HI upstream of
the shock. Recently \citet{Hidaka02} have extended the work of
\citet{Sofue95} to include spiral arm compressions, and found that HI
would be transformed into H$_2$ on a timescale of $<$ 5 Myrs upon
entering a mild spiral perturbation. After undergoing compression, the
molecular fraction reached almost unity (as in this case). After this
initial transformation, the clouds slowly returned to a mainly HI form
after about 25\,Myrs, as a result of a reduction in the overpressure
and an increase in the UV field due to star formation. The situation
in NGC\,1144 is different, because, if the work of \citet{Hidaka02} is
generally applicable to the very strong shock-wave we believe exists
in NGC\,1144, we have to explain why the molecular segregation
extended over such a large area (almost half the galaxy) rather than a
narrow region downstream of the compression.

Part of the explanation may come from the extremely rapid rotation of
NGC\,1144, combined with the peculiar streaming motions near the
stronger-than-normal shock.  In Figure 9 we show one plausible
scenario for the apparent separation of the molecular and HI
components on a galaxy-wide scale.  Gas clouds, with molecular
fractions more typical of the mid-radius molecular fraction
(i.e. mainly HI--dominated ISM) rotate rapidly in a pre-collisional
massive galaxy. The collision creates a quasi-stationary wave in the
disk through which the HI clouds travel as they move from west to east
(if the arm is trailing). The shock is strong enough to compress the
clouds, causing them to become mainly molecular down-stream of the
shock--as in the case of the galaxies discussed by
\citet{Hidaka02}. However, unlike a normal disk galaxy, which may have
circular velocities typically a 150-200 km\,s${-1}$, the clouds in
NGC\,1144 are travelling twice as fast.

Without a detailed model of the clouds entering the shock, we can only
speculate about the likely streaming motions that might
result--however the clouds are likely to be deflected inwards in a
direction tangential to the shock--perhaps creating a flow of gas
inwards towards, but overshooting the nucleus (see Figure 7). If as
much as 30\% of the rotational velocity was directed into such a flow,
then the molecular material could travel around the galaxy and arrive
roughly south of the nucleus in 30 million years--only a little longer
than the timescale envisaged by \citet{Hidaka02} for the molecular gas
to remain in the dense phase in their picture. Overpressure effects
from a pre-existing high-pressure starburst phase may also help to
extend the timescale. Eventually, these clouds would disperse through
the effects of star formation, returning unconsumed gas to a mainly
HI--dominated phase shortly thereafter.

The advantage of this admittedly simple ``sketch'' of the gas flow in
NGC\,1144, is that it seems to account for a number of previously
unexplained aspects of NGC\,1144. Firstly, the inwardly directed stream might
explain the odd fact that there are multiple velocity components in
the H$\alpha$ velocity field in the vicinity of the nucleus
\citep{Hippelein89, Keel96}, as well as large apparent gradients in
the H$\alpha$ velocity field near the dust-lane (B99). A tangential
flow along the dust-lane would also explain the strange optical loop
observed in the HST image (Figure~6a) to the south and east of the
nucleus, where the gas would swing in an anti-clockwise direction as
it interacts with ambient lower-velocity gas in that regions. This
interaction would explain why almost all of the star formation in
NGC\,1144 is concentrated either to the south and west of the nucleus
in giant extended star formation complexes. Finally, the lack of CO
emission in the center of NGC\,1144 might also be a consequence of the
fact that the molecular gas taking part in the flow would be less
dissipative than diffuse gas, and would tend to overshoot the nucleus.
Eventually, given enough time, this material may fall back to the
center. 

Figure 7 represents only one possible description of the complex phenomina
that may be taking place in NGC\,1144\footnote{We note that if the
spiral arm is leading rather than trailing (not consistent with other
observation, nor published models), the gas flow through the shock
would be reversed and the transition across the shock would be from
molecules to diffuse HI.}. Our observations emphasize the need for
gas-dynamical modeling that goes well beyond the usual single-phase
models of the galaxy interactions. Indeed, the very high velocities,
the unusually strong shock, and the likely multi-phase nature of
clouds circulating within NGC\,1144 provides a strong challenge to any
hydrodynamic modeling effort.

\section{The Nature of the Bright Double Radio Source in NGC\,1144}

What is the connection, if any, between the bright radio source and the
dust-lane in NGC\,1144?  We indicated earlier that a luminous
extra-nuclear radio source lies close to the ridge of the dust-lane.
Although the radio contours show in Figure 6b (source indicated by an arrow)
that the radio double appears blended into a single elongated source,
it is easily seen as two sources in the original 1.49\,GHz map of
\citet{Condon90}, as well as in the 8.4\,GHz VLA
observations by \citet{Kukula95}. We also observe that the northern of
the two sources is slightly elongated and less point-like in the
higher resolution 6cm VLA observations. Its combined flux is 49\,mJy
at 20cm (about 1/3 of the total 20cm flux of Arp\,118), which
corresponds to a luminosity of
L$_{\rm 20cm}=7.08\times10^{22}$ W\,Hz$^{-1}$  if at the distance of NGC\,1144\footnote{We calculated
the 1.49\,GHz absolute spectral luminosity from L$_{\rm 20cm}$=L$_{\rm
1.49GHz}=4\pi$D$^2\times$(1+z)$^{(1+\alpha)}S$, where $S$ is the
1.49\,GHz flux density and $\alpha\sim0.9$ \citep{Condon90}.}. To put
this in perspective, it is similar
to the {\em total} 20cm luminosity of M\,82 and only $\sim$3.7 times
less than that of  Arp\,220 \citep{Condon90}. Its steep spectral
index \citep[$\sim$0.9 see][]{Jeske86} suggests a non-thermal origin.

Aside for the
coincidence of the source lying in a ``hole'' in the CO distribution,
the emission is unlikley to be a background galaxy on statistical
grounds. According to J. Condon (NRAO -- Private Communication) the
probability of finding a 50\,mJy source so close to the nucleus of
NGC\,1144 is about 1 part in 10,000 and consequently a spatial
connection with Arp\,118 is implied.
 
There are a number of possible explanations for the formation of a
50mJy radio source in the inner disk of NGC\,1144.  We can rule out
radio supernovae (RSNe) for the radio emission. Firstly the spectrum
of RSNe is much flatter ($\le 0.3$) than measured values for the
source \citep{Jeske86}. Our continuum observations (obtained as part
of the data cube presented in this paper) recover the same flux for
the double source as previously reported from observations obtained in
1987--88 \citep{Jeske86, Condon90}. The fact that the source has not
faded by a measurable amount over at least 14 years would contradict
our current understanding of RSNe that they fade rapidly after
reaching a peak of L$_{\rm 6cm}\sim$ 2 $\times$
10$^{20}$\,W\,Hz$^{-1}$
\citep[see][]{Chevalier82,Weiler98}. Furthermore, the luminosity of
the double taken together is significantly brighter than the brightest
known RSNe in M\,82 or Arp\,220. Since the peak 1.67\,GHz radio power
for the brightest RSN detected by \citet{Smith98} in Arp\,220 is
1.4$\times$10$^{21}$\,W\,Hz$^{-1}$, then one would need to have more
than 45 {\em currently active} RSNe to account for the radio continuum
flux. Such a scenario appears unlikely.

Another more likely explanation for the radio source is that it is
due to prompt star formation from an unusually large compressed cloud
which in turn generated multiple  supernova explosions. The shock
from the SNe could have destroyed/dissociated the molecules, apparently
leaving the HI unaffected, and created a series of unresolved
bubble-like emission regions similar to giant supernova remnants. The
supernova rate needed to generate the observed radio flux can be
calculated using the definition of \citet{Condon90b}\footnote{The
supernova rate can be estimated from the formula \rm L$_{\rm NT}$(\rm
W\,\rm Hz$^{-1}$) = 1.3$\times$10$^{23}$ ($\frac{\nu}{1\,\rm
GHz})^{-\alpha}$ ($\frac{\rm R_{\rm SN}}{\rm 1\,yr}$), where L$_{\rm
NT}$\,=\,L$_{\rm 20cm}=7.08\times10^{22}$ W\,Hz$^{-1}$ is the
non-thermal radio luminosity, $\alpha$=0.9 is the radio spectral
index, and R$_{\rm SN}$ is the rate of type II supernova
\citep{Condon90b}.}  and is found to be R$_{\rm
SN}$=0.78\,yr$^{-1}$. This rate is surprisingly high for a region of
just $\sim$1\,kpc in size, since it is $\sim$7 times higher than the
total supernovae rate of M\,82 \citep{Huang94}, and similar to what is
found in nuclei of ULIRGs. What is less clear
is why this, and only this part of the shock-wave experienced such
enhanced activity.

Based on numerical models, \citet{Lamb98} speculated that the source
marks the point of impact of the companion galaxy which collided
22\,Myr ago. Star formation, created at that point by the increased
density at the impact time, may have created many supernova which
would provide the energy for the radio emission. One thing that is not
clear in their models is why the impact site should lie so close to
the current spiral shock. At the elapse-time in their simulation when the
best match is made with the morphology of NGC\,1144, the spiral-shock
has already propagated well beyond the impact site and would no longer
be in contact. Further observations will be required
to gain more understanding of the nature of the radio-source and its
environment.

\section{Conclusions}

We present high--resolution (VLA B \& C array) HI observations of the
Arp\,118 (NGC 1144/3) system. Our main conclusions are:

1) The warmer neutral hydrogen gas and the cooler molecular gas is
   largely segregated azimuthally around NGC\,1144 (leading to a
   Yin-Yang like-separation of the gaseous ISM).  Most of the
   molecular gas is concentrated in the south-eastern half of the galaxy,
   and the HI in the north-west. The HI and molecular gas track the
   coherent, but very rapidly rotating, H$\alpha$ disk in NGC\,1144.
   Because the two components appear segregated spatially, the gas is
   also kinematically segregated--the HI gas largely samples the low
   radial velocity end of the rotating disk, while the molecules
   sample the high-velocites. A complete picture
   of the rotation of the gas can only be seen when both
   components are considered together.

2) We suggest that the collision between NGC\,1143 and NGC\,1144 has
   initiated a phase transition in the ISM as outlined by
   \citet{Elmegreen93}.  HI and CO clouds are seen along an interface
   region which coincides with a dark, extremely sharp, dust-lane seen
   on an HST image of the galaxy. Combined with other evidence, we
   suggest that the dust-lane is a collisionally induced shock-wave
   that may be responsible for the conversion of diffuse HI clouds
   (molecular fraction $<$~0.50) into dense molecular-dominated
   (molecular fraction $>$~0.96) cloud complexes as they pass through
   the shock-wave. Because of the extremely rapid rotational
   velocities in the galaxy (the overall velocity spread is 1200 km
   s$^{-1}$), the gas can orbit a significant distance around the
   galaxy after the compression before the residual molecular clouds
   (not consumed in the associated star formation) are converted back
   to HI again--leading to an apparent large-scale segregation.

3) Further evidence of a strong shock within the disk of NGC\,1144
   comes from the discovery of two HI absorption-lines, separated by
   185 km s$^{-1}$, which are seen against the brightest radio source which
   lies close to the center-line of the dust lane.  
   The HI absorption lines lie on either side of the
   dust-lane, and may be sampling gas clouds entering the shock, and
   reforming with much lower column-density further down-stream.  If
   this explanation is correct, then the velocity-jump implies large
   streaming motions almost parallel to the dust-feature which would
   funnel gas inwards to overshoot the nucleus. This is consistent
   with both the loopy morphology of NGC\,1144 in this region, and the
   unusually large non--nuclear star formation in
   the disk. This high angular--momentum gas stream may explain the
   lack of molecular pile-up in the center of the galaxy--a situation
   that is unusual for a violently collisional system with this level
   of far-IR luminosity (L$_{\rm IR}$ = 2.2$\times$10$^{11}$\,L$_{\sun}$) .

4) Although faint (2--3 mJy) radio emission is seen associated with
   the dust-lane, symptomatic of mild magnetic field compression, the
   double radio source mentioned above may be a hot-spot
   20--30$\times$ more powerful than the rest of the
   dust-lane. Although we cannot completely rule out the possibility
   that the source is a distant background object, the positional
   agreement with the dust-lane center suggests that it is associated
   with NGC\,1144. We speculate that the source represents emission
   from cosmic rays released by an episode of massive star formation
   and supernova explosions. A gap is also seen in the molecular cloud
   distribution coincident with the radio source position, suggesting
   that the process that created the source has either disrupted the
   transformation of HI into H$_2$ in that region, or has rapidly
   turned the H$_2$ into massive stars.  Further observations,
   especially in the near and mid-IR will be needed to understand the
   processes which have given rise to the bright radio source region.

5) HI emission is observed extending beyond the outer disk of NG\,1144
   towards the elliptical companion NGC\,1143. This  gas 
   may represent the early stages of matter transfer between the two
   galaxies. The gas mass involved is small (M$_{\rm HI} = 2 \times 10^{8}$
   M$_{\odot}$) but provides further evidence that the two galaxies
   have recently collided.

\acknowledgments

PNA wishes to thank W. Reach (SSC-Caltech) for interesting
discussions, and James Condon (NRAO) for the radio continuum image of
Arp\,118 and discussions relating to the eastern radio double source.
VC would like to acknowledge the partial support of JPL contract
960803.  This research has made use of the NASA/IPAC Extragalactic
Database (NED) which is operated by the Jet Propulsion Laboratory,
California Institute of Technology, under contract with the National
Aeronautics and Space Administration. Based on observations made with
the NASA/ESA Hubble Space Telescope, obtained from data archive at the
Space Telescope Science Institute. STScI is operated by the
Association of Universities for Research in Astronomy, Inc. under the
NASA contract NAS 5-26555.

\clearpage






\clearpage

\begin{deluxetable}{lcccccccc}
\tablecaption{Properties of the Galaxies\label{tbl_1}}
\tablewidth{0pc}
\startdata \\
\tableline 
\tableline \\

Name & Type & D &V$_{helio}$\tablenotemark{a} &
$\Delta$V$_{\rm HI}$ &
$\Delta$V$_{\rm CO+HI}$\tablenotemark{b} &
M$_{\rm HI}$/M$_\sun$\tablenotemark{c} & M$_{\rm HI}$/M$_\sun$\tablenotemark{d} & M$_{\rm HI}$/M$_\sun$\tablenotemark{e}
\\

 & & (Mpc) & (km\,s$^{-1}$) & (km\,s$^{-1}$) & (km\,s$^{-1}$) & ($\times$~10$^{9}$) & ($\times$~10$^{9}$) & ($\times$~10$^{9}$) \\   
\tableline \\

NGC\,1144 & S$_{pec}$--Ring & 110 & 8648$\pm$14 & 170\tablenotemark{b} & 1120 & 2.0 & 5.6 & 7.0 \\

NGC\,1143 & E/S0$_{pec}$ & 110 & 8459$\pm$30 & 44\tablenotemark{f} & -- & 0.2\tablenotemark{f} & -- & -- \\


\tablenotetext{a}{The optical velocities are from \citet{Keel96}.}
\tablenotetext{b}{from our earlier paper (B99).}
\tablenotetext{c}{HI masses from the brighter features from this paper based on 8 $\times$ 8 arcsecs restored beam (see text).}
\tablenotetext{d}{HI masses from this paper based on 15 $\times$ 15 arcsecs beam (see text). Note that considerably more extended emission is detected.}
\tablenotetext{e}{HI properties from low-resolution 21.2 $\times$ 17.0 arcsecs beam  presented in B99 using C-array only.}
\tablenotetext{f}{Assuming extension at V = 8270\,km\,s$^{-1}$ within optical envelope is associated with NGC\,1143 (see text). }
\enddata
\end{deluxetable}

\begin{deluxetable}{lccccc}
\tablecaption{Properties of Absorption--Line Systems\label{tbl_2}}
\tablewidth{0pc}
\startdata \\
\tableline 
\tableline \\

System & V$_{helio}$ & $\Delta$V$_{\rm FWHM}$ & $<$ $\tau$
$>$\tablenotemark{a} & $\tau$$_{max}$ & N$_{\rm HI}$ \\ & (km\,s$^{-1}$) & (km\,s$^{-1}$) &
& & $\times$~10$^{20}$ atoms cm$^{-2}$ \\
\tableline \\
System I & 9050$\pm$20 & 100$\pm$30 & 0.023 & 0.027 &  5.5$\times$ T$_{s100}$ \\
System II & 8825$\pm$21 & 140$\pm$30 & 0.036 & 0.044 & 11.5$\times$ T$_{s100}$ \\   

\tableline \\


\tablenotetext{a}{Optical depths calculated using background continuum
map derived from this data--not shown, and assuming the radio source
fills the 6.2~$\times$~6.3 arcsec beam (this is therefore a strict
lower limit--see text). The average optical depths shown are the
average optical depths over the channels in which absorption was
detected. Note T$_{S100}$ = an assumed spin temperature in units of 100 K.} \enddata
\end{deluxetable}

\newpage

\clearpage 
\begin{figure} 
\plotone{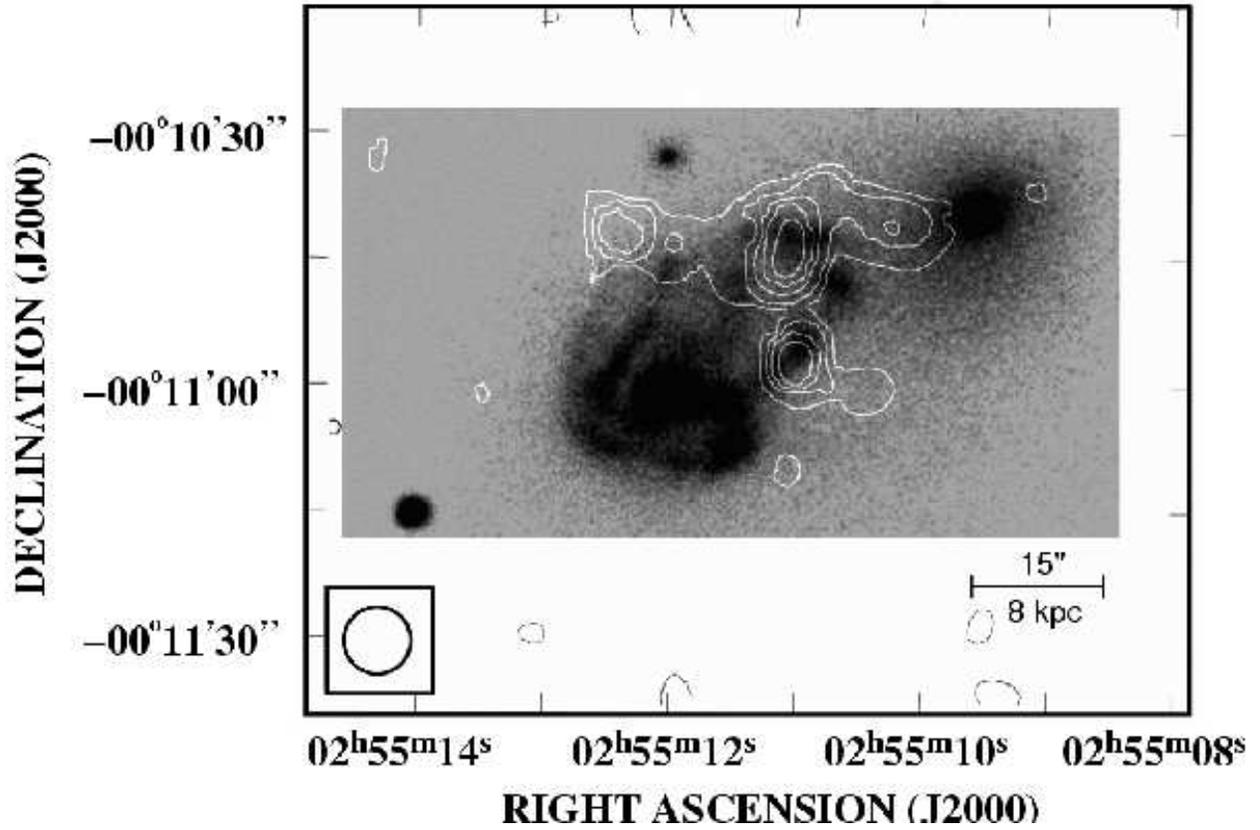}
\figurenum{1} 
\caption{The integrated HI surface density map superimposed on an
r-band plate of Arp\,118 from \citet{Hippelein89}. The ring galaxy NGC\,1144 is to the south-east, and E2/S0 galaxy NGC\,1143 to the north-west. 
Contour levels are 1, 2, 3, 4, 5, 10, 15 $\times$ 33.9\,mJy beam$^{-1}$ km\,s$^{-1}$ or units of 5.8\,$\times$ 10$^{20}$ atoms cm$^{-2}$.}
\end{figure} 

\begin{figure} 
\plotone{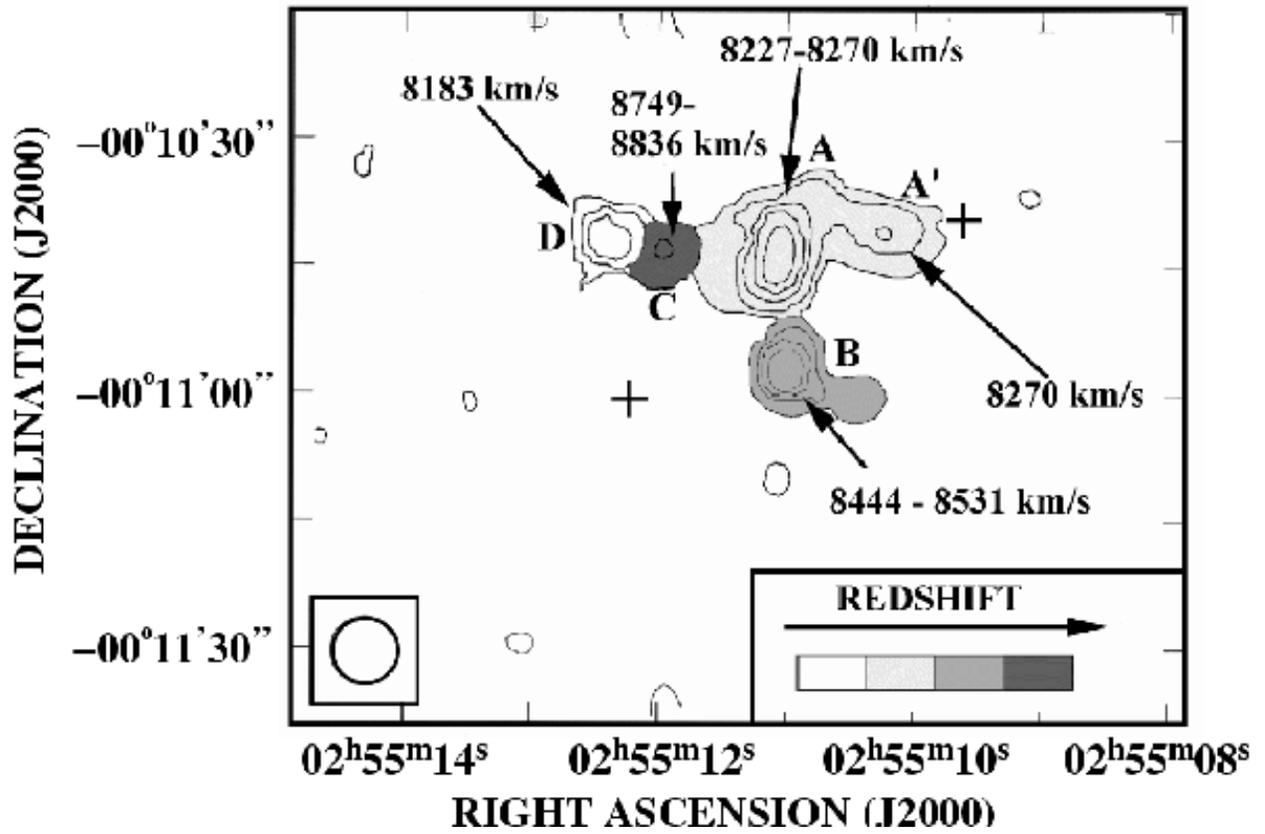}
\figurenum{2} 
\caption{The heliocentric radial velocities of the HI emission regions~
(km\,s$^{-1}$). The crosses mark the centers of NGC\,1144 and NGC\,1143.}
\end{figure} 

\begin{figure} 
\plotone{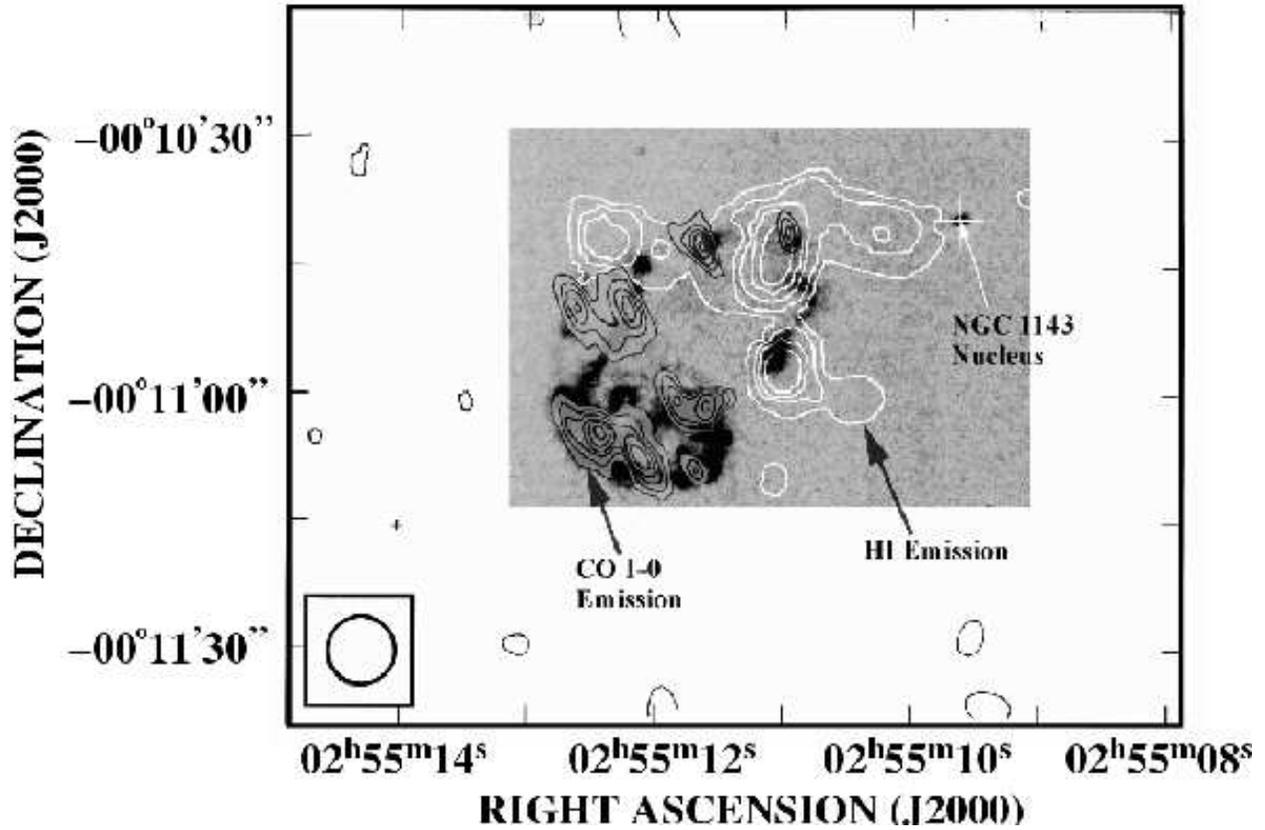}
\figurenum{3} 
\caption{ The combined molecular and neutral gas content of Arp
118 superimposed on the greayscale image of H$\alpha$ emission from \citet{Hippelein89}. The $^{12}$CO(1-0) emission contours from \citet{Gao97} are shown
in grey starting at 5.14\,Jy\,km\,s$^{-1}$\,beam$^{-1}$ and increasing
by 3.86\,Jy\,km\,s$^{-1}$\,beam$^{-1}$. The HI contours of Figure 1
are also included in white--(for units see Figure 1). Note the
remarkable segregation of the neutral and molecular gas.}
\end{figure}

\begin{figure}
\plotone{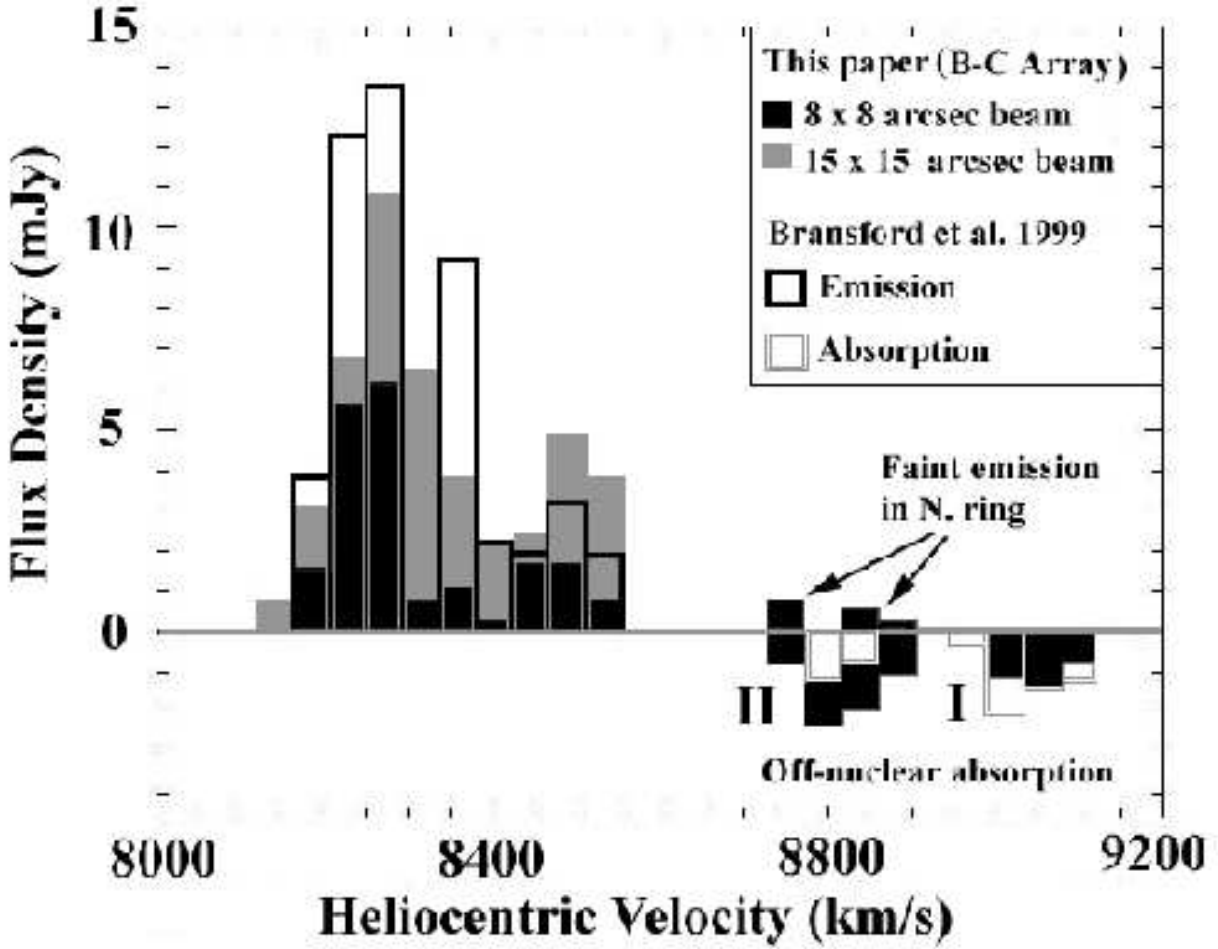}
\figurenum{4}
\caption{ The integrated HI spectrum derived from the new combined B-C
Array data and our original lower-resolution C-array
observations (see key for color coding). Newly detected weak emission near
V~=~8800\,km\,s$^{-1}$ from the northern ring was absent from the older
observations because it was nullified by the faint absorption seen
near the center of the galaxy.  Despite this, the new observations
confirm the earlier result that the HI emission occupies a much smaller
range of radial velocities than seen in the ionized gas disk because of spatial segregation.}
\end{figure}

\begin{figure} 
\plotone{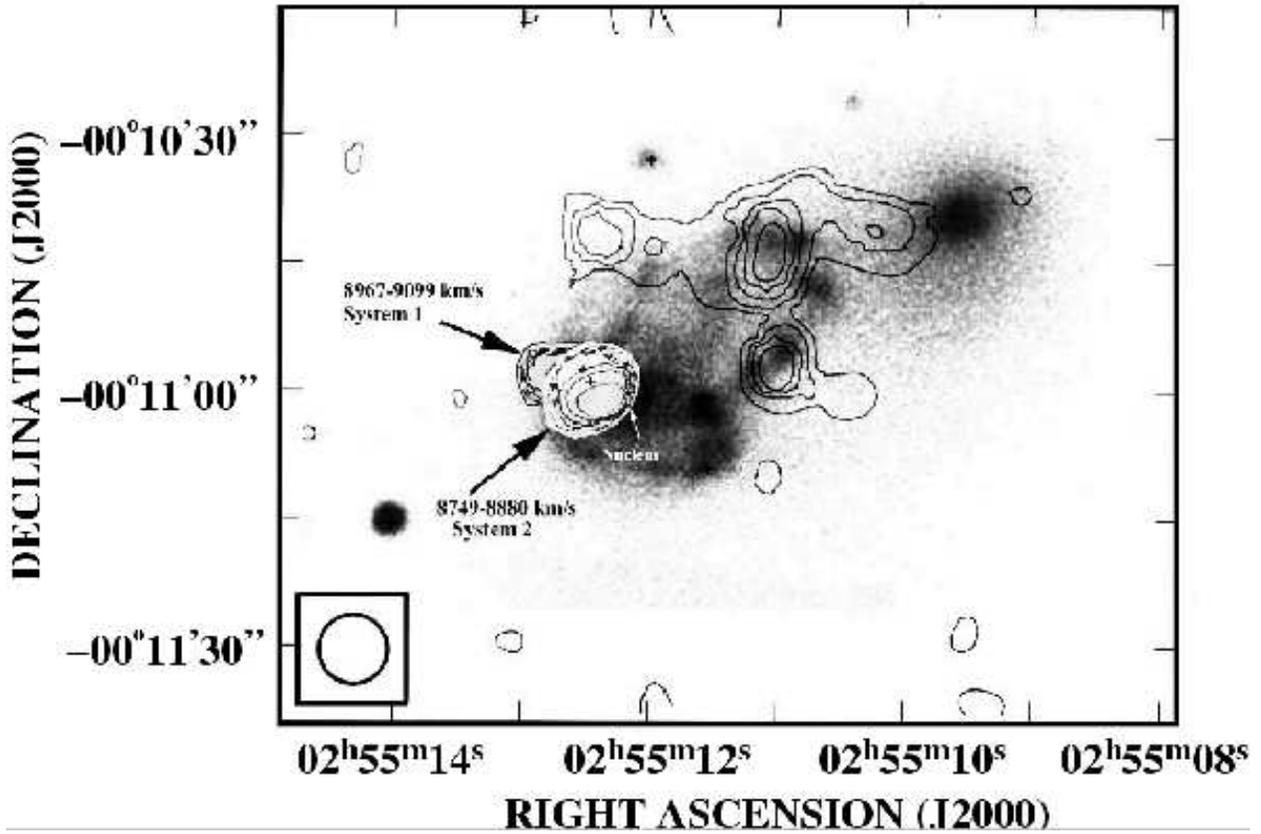}
\figurenum{5} 
\caption{The HI absorption (System 1 and 2) seen superimposed on the
R-band image of Arp 118. Note that the absorption is seen against the
50 mJy double radio source (dark crosses denote centroid of double
source) discovered by \citet{Condon90} which lies just to the east of
the nucleus.  The fainter nucleus of NGC\,1144 does not show HI
absorption (white cross). Absorption contour units are $-$3, $-$2.7, $-$2.3,
$-$1.9, $-$1.5\,mJy~beam$^{-1}$. Emission lines contours in black--units
see Figure 1.}
\end{figure}

\begin{figure}
\plotone{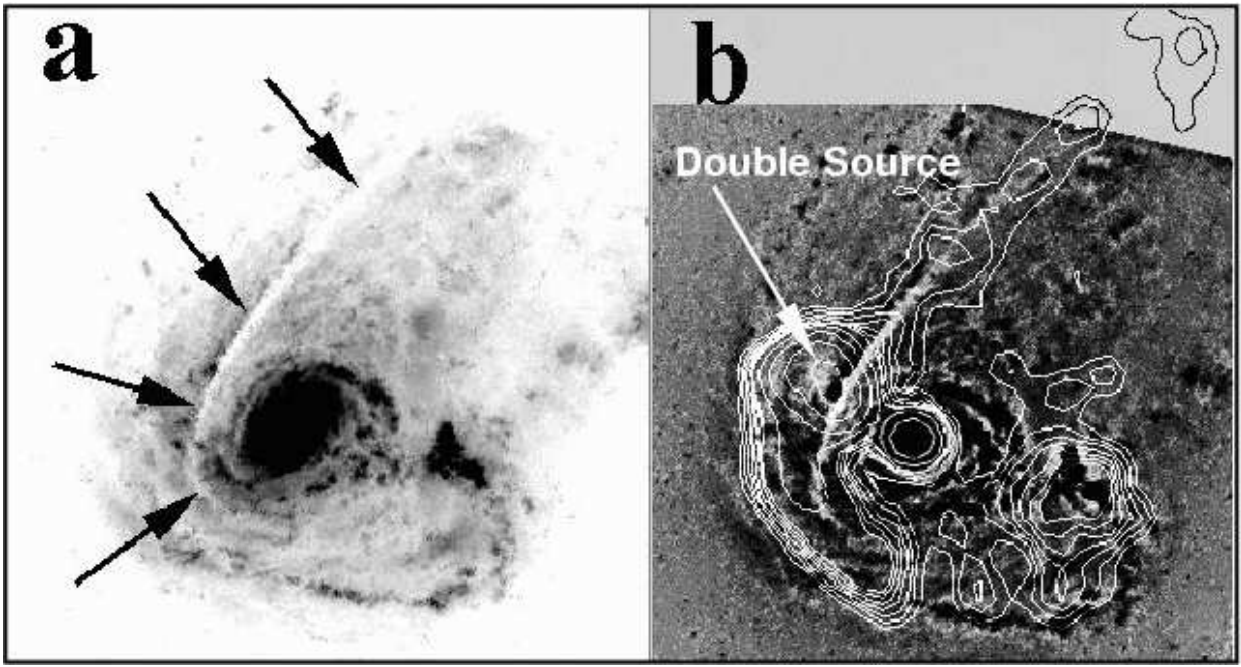}
\figurenum{6}
\caption{The F606 red continuum image from WFPC2: (a)(grayscale)
showing the dominant morphological features of NGC\,1144. Note the
prominent dust lane (arrows) which is seen more clearly in (b), where
the image (grayscale) has been unsharp masked to reveal the
small-scale structure in the image. Contours show 20cm VLA radio
continuum emission from \citet{Condon90}, and seems to follow the dust
lane very closely over most of its length.
}
\end{figure}

\begin{figure} 
\plotone{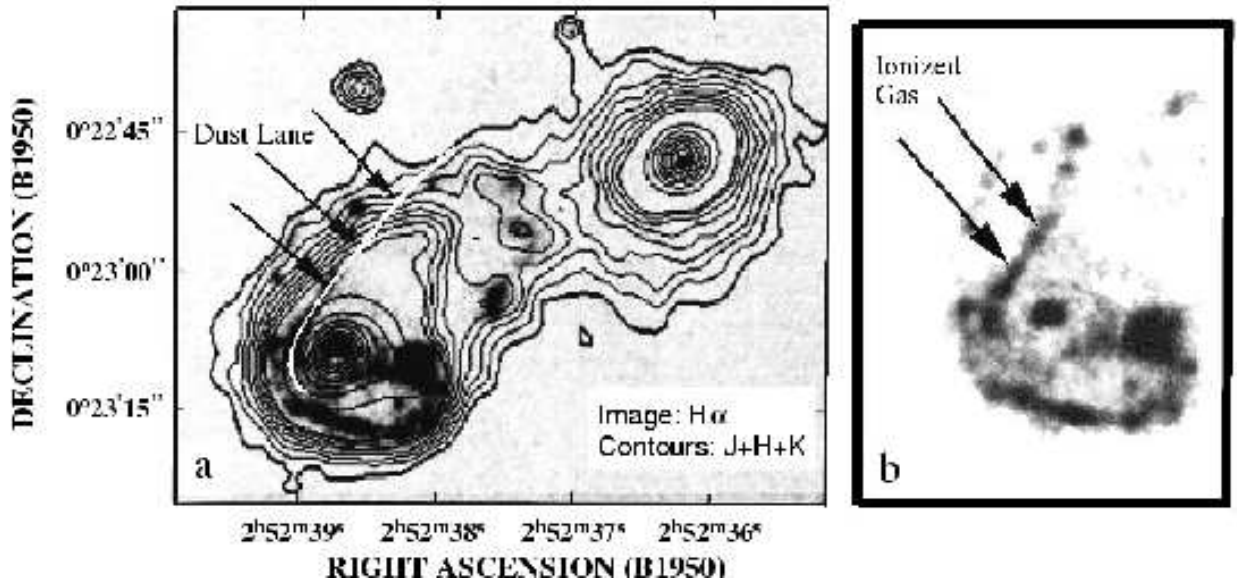}
\figurenum{7} 

\caption{(a) The sum of J, H and K-band images of Arp 118 from 2MASS
(contours) superimposed on the H$\alpha$+N[II] image (grayscale). The
dust lane is also marked as a white line. (b) For clarity, the inner
part of NGC\,1144 is shown again in H$\alpha$+[NII] light without the
dust lane position. The bright ionized gas filament marked with the
arrow is almost coincident with the inner segment of the
dust-lane. Note that the dust-lane/H$\alpha$ filament shows a complete
mis-match with the near-IR light, implying that the feature is a
transient hydrodynamic structure rather than a density wave with an
underlying old stellar population.}

\end{figure}

\begin{figure} 
\plotone{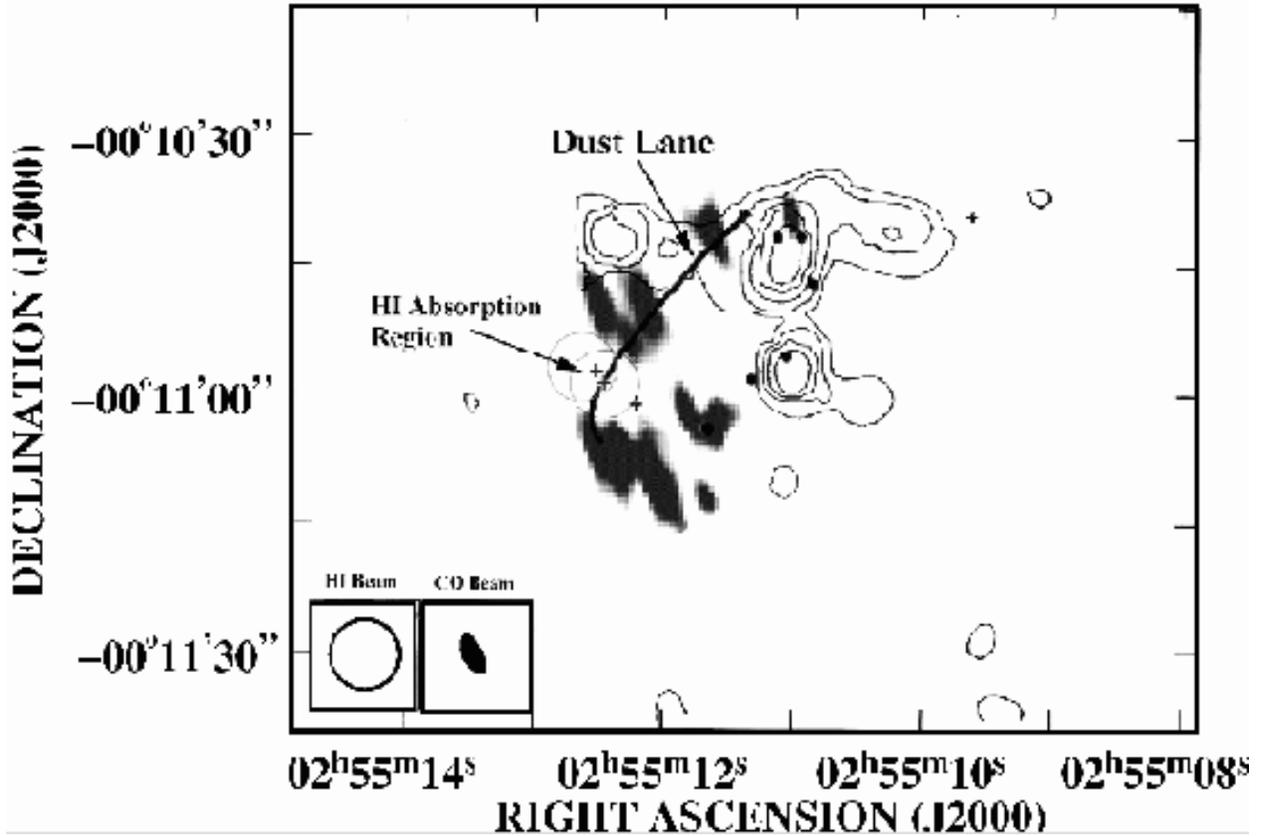}
\figurenum{8} 

\caption{The locus of the center of dust lane marked onto a plot of
the HI and CO distributions in Arp 118. The dust lane's position was
traced a little beyond the HST WFPC2 image of Figure 5, by assuming
that it is traced by the radio continuum ridge. The three crosses mark
the positions of the double radio source (close to the dust lane), and
the nucleus to the south and west.}
\end{figure}

\newpage

\begin{figure}[p] 
\epsscale{0.5}\plotone{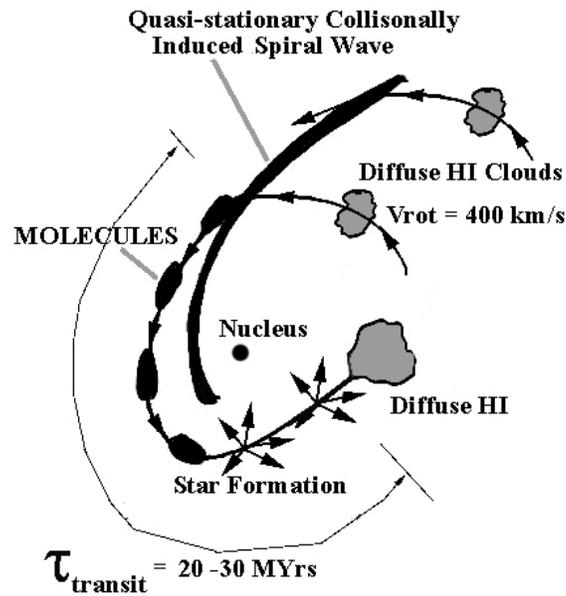}
\figurenum{9} 
\caption{A schematic representation of how HI and molecules may be
segregated in NGC\,1144. HI clouds, circulating in a counter-clockwise
manner in the rotating disk of NGC 1144, encounter the strong shock
wave are compressed becoming mainly molecular \citep{Elmegreen93}.
The clouds are deflected by the shock inwards.  Eventually, star
formation destroys the molecules and they circulate rapidly to the
west, becoming mainly neutral hydrogen-dominated again. The unusually
rapid rotation of NGC 1144, combined with the transient one-armed 
spiral shock-wave leads to the appearence of
large-scale segregation of the HI and molecules.}

\end{figure} 
\end{document}